%% file: sample-authordraft.tex
\documentclass[sigconf]{acmart}

\AtBeginDocument{%
  \providecommand\BibTeX{{%
    \normalfont B\kern-0.5em{\scshape i\kern-0.25em b}\kern-0.8em\TeX}}}

\setcopyright{acmcopyright}
\copyrightyear{2018}
\acmYear{2018}
\acmDOI{10.1145/1122445.1122456}

\acmConference[ICAIF-2020]{2020 ACM International Conference on AI in Finance}{October 15--16, 2020}{NY}
\acmBooktitle{2020 ACM International Conference on AI in Finance,
  October 15--16, 2020, NY}
\acmPrice{15.00}
\acmISBN{978-1-4503-XXXX-X/18/06}

\include{homemade_macros}

\include{mymacros_2}

\newtheorem{remark}{Remark}

\begin{document}

\title{Sig-SDEs model for quantitative finance. }

\author{Imanol Perez Arribas}
\email{imanol.perez@maths.ox.ac.uk}
\affiliation{%
  \institution{University of Oxford}
  \institution{Alan Turing Institute}
  \country{United Kingdom}
}

\author{Cristopher Salvi}
\email{cristopher.salvi@maths.ox.ac.uk}
\affiliation{%
  \institution{University of Oxford}
  \institution{Alan Turing Institute}
  \country{United Kingdom}
}

\author{Lukasz Szpruch}
\email{l.szpruch@ed.ac.uk}
\affiliation{%
  \institution{University of Edinburgh}
  \institution{Alan Turing Institute}
  \country{United Kingdom}
}

\renewcommand{\shortauthors}{Perez Arribas, Salvi and Szpruch}

\begin{abstract}
Mathematical models, calibrated to data,  have become ubiquitous to make key decision processes in modern quantitative finance. In this work, we propose a novel framework for data-driven  model selection by integrating a classical quantitative setup with a generative modelling approach. Leveraging the properties of the signature, a well-known path-transform from stochastic analysis that recently emerged as leading machine learning technology for learning time-series data, we develop the Sig-SDE model.  Sig-SDE provides a new perspective on neural SDEs and can be calibrated to exotic financial products that depend, in a non-linear way, on the whole trajectory of asset prices. Furthermore, we our approach enables to consistently calibrate under the pricing measure $\mathbb Q$  and real-world measure $\mathbb P$. Finally, we demonstrate the ability of Sig-SDE to simulate future possible market scenarios needed for computing risk profiles or hedging strategies. Importantly, this new model is underpinned by rigorous mathematical analysis, that under appropriate conditions provides theoretical guarantees for convergence of the presented algorithms. 
\end{abstract}




\maketitle

\setlength\parindent{0pt}

\section{Introduction}

The question of finding a parsimonious model that well represents empirical data has been of paramount importance in quantitative finance.  The modelling choice is dictated by the desire to fit and explain the available data, but is also subject to computational considerations.  Inevitably, all models can only provide an approximation to reality, and the risk of using inadequate ones is hard to detect. 
A classical approach consists in fixing a class of parametric models, with a number of parameters that is significantly smaller than the number of available data points. Next, in the process called calibration, the goal is to solve a data-dependent optimization problem yielding an optimal choice of model parameters. The main challenge, of course, is to decide what class of models one should choose from. The theory of statistical learning \cite{vapnik2013nature} tell us that to simple models cannot fit the data, and to complex one are not expected to generalise to unseen observations. In modern machine learning approaches, one usually starts by defining a  highly oveparametrised model from some universality class, exhibiting a number of parameters often exceeding the number of data points, and let (stochastic) gradient algorithms find the best configuration of parameters yielding a calibrated model. In this work, we find a middle ground between the two approaches. We develop a new framework for systematic model selection that exhibits universal approximation properties, and we provide a explicit solution to the optimization used in its calibration, that completely removes the need to deploy expensive gradient descent algorithms. Importantly the class of models that we consider builds upon classical risk models that are well underpinned by research on quantitative finance. 

The mathematical object at the core of this work is the expected signature of a path, whose properties are well-understood in the field of stochastic analysis. It allows to identify a linear structure underpinning the high non-linearity of the sequential data we work with. This linear structure leads to a massive speed-up of calibration, pricing, and generation of future scenarios. Our approach provides a new systematic model selection mechanism, that can also be deployed to calibrate classical non-Markovian models in a computationally efficient way. Signatures have been deployed to solve various tasks in mathematical finance, such as options pricing and hedging \cite{lyons2019nonparametric, lyons2020numerical}, high frequency optimal execution \cite{kalsi2020optimal, cartea2020optimal} and others \cite{lyons2014feature, gyurko2013extracting}. They have also been applied in several areas of machine learning \cite{kidger2019deep, yang2015chinese, yang2016deepwriterid, yang2016dropsample, yang2016rotation, yang2017leveraging, xie2017learning, li2017lpsnet, chevyrev2018signature, kiraly2016kernels}.

\subsection{Sig-SDE Model}
    
Let $X:[0, T] \to \mathbb R^d$ denote the price process of an arbitrary financial asset under the pricing measure $\mathbb{Q}$. To ensure the no-arbitrage assumption is not violated, $X$ typically is given by the solution of the following \textit{Stochastic Differential Equation} (SDE)
\begin{equation}\label{eq nsde}
    dX_t = \Sigma_t dW_t, \quad X_0=x\,,
\end{equation}
where $W$ is a one-dimensional Brownian motion and $\Sigma_t$ is an adapted process (the volatility process). Model \eqref{eq nsde} accommodates many standard risk models used e.g: the classical Black--Scholes model assumes that volatility is proportional to the spot price, i.e. $\Sigma_t := \sigma X_t$ with $\sigma\in \mathbb R$ constant; the local volatility model assumes that $\Sigma_t := \sigma(t, X_t) X_t$, where $\sigma(\cdot, \cdot)$ (called local volatility surface) depends on both time and spot. Hence, it is a generalisation of the Black--Scholes model;
 various stochastic volatility model assume that $\Sigma_t := \sigma_t X_t$ with $\sigma_t^2$ following some diffusion process;
 the SABR model chooses $\Sigma_t := \sigma_t X_t^\beta$, with $\beta\in [0, 1]$ and where $\sigma_t$ follows a diffusion process.

A natural question would be whether one can find a model for the volatility process $\Sigma_t$ that is large enough to include all the classical models such as the ones mentioned above and that would allow for systematic a data driven model selection. We will require such a model to satisfy the following requirements:

\begin{enumerate}
\item \textbf{Universality}. The model should be able to approximate arbitrarily well the dynamics of classical models.
\item \textbf{Efficient calibration}. Given market prices for a family of options, it should be possible to efficiently calibrate the model so that it correctly prices the family of options.
\item \textbf{Fast pricing}. Ideally, it should be possible to quickly price (potentially exotic) options under the model without using Monte Carlo techniques.
\item \textbf{Efficient simulation}. Sampling trajectories from the model should be computationally cheap and efficient.
\end{enumerate}

An example of a model that satisfies point 1. above is a \textit{neural network model}, where the volatility process $\Sigma_t$ is approximated by a neural network $\mathcal{NN}^\theta(t, (W_s)_{s\in[0,t]})$ with parameters $\theta$. Such a model would be able to approximate a rich class of classical models. However, the calibration and pricing of such models would involve performing multiple Monte Carlo simulations on each epoch, which might be expensive if done naively. See however, \cite{cuchiero2020generative,Szpruch2020}. 

The aim of this paper is to propose a model for asset price dynamics that, we believe, satisfies all four points above. Our technique models the volatility process $\Sigma_t$ as
\begin{equation}
\Sigma_t = \langle \ell^{N}, \widehat{\mathbb W}_{0,t}\rangle
\end{equation}
where $\ell^{N}$ is the model parameters and $\widehat{\mathbb W}_{0,t}$ is the \textit{signature} (c.f definition \ref{def:sig}) of the stochastic process $\widehat W_t := (t, W_t)$. The motivation for choosing the signature as the main building block of this paper is anchored in a very powerful result for universal approximation of functions based on the celebrated \textit{Stone-Weierstrass Theorem} that we present next in an informal manner (for more technical details see \cite[Proposition 3]{fermanian2019embedding})

\begin{theorem}
Consider a compact set $K$ of continuous $\mathbb{R}^d$-valued paths. Denote by $S$ the function that maps a path $X$ from $K$ to its signature $\mathbb{X}$. Let $f:K \to \mathbb{R}$ be any any continuous functions. Then, for any $\epsilon>0$ and any path $X \in K$, there exists a linear function $l^\infty$ acting on the signature such that
\begin{equation}
    ||f(X) - \langle l^\infty, \mathbb{X}\rangle||_\infty < \epsilon
\end{equation}
\end{theorem}

In other words, any continuous function on a compact set of paths can be uniformly well approximated by a linear combination of terms of the signature. This universal approximation property is similar to the one provided by Neural Networks (NN). However, as we will discuss below, NN models depend on a very large collection of parameters that need to be optimized via expensive back-propagation-based techniques, whilst the optimization needed in our Sig-SDE model consists of a simple linear regression on the terms of the signature. In this way, the signature can be thought of as a feature map for paths that provides a linear basis for the space of continuous functions on paths. In the setting of SDEs, sample paths are Brownian and solutions are images of these sample trajectories by a continuous functions that one wishes to approximate from a set of observations. Our Sig-SDE model will rely upon the universality of the signature to approximate such functions acting on Brownian trajectories. Importantly, the signature of a realisation of a semimartingale provides a unique representation of the sample trajectory \cite{hambly2010uniqueness, boedihardjo2016signature}. Similarly, the \textit{expected signature} -- i.e. the collection of the expectations of the iterated integrals -- provides a unique representation of the law of the semimartingale \cite{chevyrev2016characteristic}.

Note that model calibration is an example of generative modelling \cite{goodfellow2014generative,kingma2013auto}. 
Indeed, recall that if one knew prices of traded liquid derivatives, then one can approximate the pricing measure from market data \cite{breeden1978prices,lyons2019nonparametric}. We denote this measure by $\mathbb Q^{real}$. 

We know that when equation \eqref{eq nsde} admits a strong solution then there exists a measurable map $G:\mathbb R \times C([0,T]) \rightarrow C([0,T])$ such that
\begin{equation}
    X = G(x, (W_s)_{s\in[0,T]})
\end{equation}
as shown in \cite[Corollary 3.23]{karatzas2012brownian}. If $G_t$ denotes the projection of $G$ given by $X_t:=G_t(\xi, (W_s)_{s\in[0,t]})$, then one can view \eqref{eq nsde} as a generative model that maps  $\mu_0$ supported on $\mathbb R^d$ into $(G_t)_{\#}\mu_0=Q_t^{\theta}$. Note that by construction $G$ is a casual transport map i.e a transport map that is adapted to the filtration $\mathcal F_t$ \cite{acciaio2019causal}. In practice, one is interested in finding such a transport map from a family of parametrised functions $G^\theta$. One then looks for a $\theta$ such that $G^\theta_{\#}\mu_0$ is a good approximation of $Q^{real}$ with respect to a metric specified by the user. In this paper the family of transport maps $G^\theta$ is given by linear functions on signatures (or linear functionals below).

\section{Notation and preliminaries}
We begin by introducing some notation and preliminary results that are used in this paper.

\subsection{Multi-indices}

\begin{definition}
Let $d \in \mathbb{N}$. For any $n\geq 0$, we call an $n$-dimensional \textit{$d$-multi-index} any $n$-tuple of non-negative integers of  the form $K = (k_1, \ldots, k_n)$ such that $k_i \in \{1,\ldots,d\}$ for all $i\in \{1,\ldots, n\}$. We denote its length by $|K|=n$. The empty multi-index is denoted by $\o$. We denote by $\mathcal I_d$ the set of all $d$-multi-indices, and by $\mathcal{I}^n_d \subset \mathcal{I}_d$ the set of all $d$-multi-indices of length at most $n\in \mathbb N$.
\end{definition}

\begin{definition}[Concatenation of multi-indices]
Let $I = (i_1, \ldots, i_p)$ and  $J = (j_1, \ldots, j_q)$ be any two multi-indices in $\mathcal{I}_d$. Their \textit{concatenation product} $\otimes$ as the multi-index $I \otimes J = (i_1, \ldots, i_m, j_1, \ldots, j_n) \in \mathcal{I}_d$.
\end{definition}

\begin{example}$\phantom{ }$
\begin{enumerate}
    \item $(1, 3) \otimes (2, 2) = (1, 3, 2, 2)$.
    \item $(2, 1, 3) \otimes (1) = (2, 1, 3, 1)$.
    \item $(2, 2) \otimes \o = (2, 2)$.
\end{enumerate}
\end{example}

\subsection{Linear functionals}

\begin{definition}[Linear functional]
For a given $d\geq 1$, a \textit{linear functional} is a (possibly infinite) sequence of real numbers indexed by multi-indices in $\mathcal{I}_d$ of the following form
\begin{equation}
    F=\{F^{(K)} \in \mathbb{R} : K \in \mathcal{I}_d\}.
\end{equation}
\end{definition}

We note that a multi-index $K \in \mathcal{I}_d$ is always a linear functional. Both concatenation $\otimes$ and $\shuffle$ can be extended by linearity to operations on linear functionals. We will now define two basic operations on linear functionals that will be used throughout the paper.

\begin{definition} For any two linear functionals $F, G$ and any real numbers $\alpha, \beta \in \mathbb{R}$ define
\begin{equation}
    \alpha F + \beta G = \{\alpha F^{(K)} + \beta G^{(K)} \in \mathbb{R}: K \in \mathcal{I}_d\}
\end{equation} 
and
\begin{equation}
    \langle F, G \rangle = \sum_{K \in \mathcal{I}_d} F^{(K)} G^{(K)} \in \mathbb R
\end{equation}
\end{definition}

\subsection{Signatures}\label{sec:signatures}

\textit{Rough paths} theory can be briefly described as a non-linear extension of the classical theory of controlled differential equations which is robust enough to allow a deterministic treatment of stochastic differential equations controlled by much rougher signals than semi-martingales \cite{lyons2007differential}.

\begin{definition}[Signature]\label{def:sig}
Let $X:[0, T] \to \mathbb R^d$ be a continuous semimartingale. The Signature of $X$ over a time interval $[s, t]\subset  [0, T]$ is the linear functional $\mathbb X_{s, t} := \{\mathbb X_{s, t}^{(K)} \in \mathbb{R} : K \in \mathcal{I}_d\}$, such that $\mathbb{X}_{s,t}^{(\o)} = 1$ and so that for any $n\geq1$  and $K = \hat{K}\otimes a \in \mathcal{I}^n_d$, with $a \in \{1, \ldots, d\}$ and $\hat{K} \in \mathcal{I}^{n-1}_d$ we have 
\begin{equation}\label{eq:sig}
    \mathbb X_{s, t}^{(K)} = \int_s^t \mathbb X^{(\hat{K})}_{s, u} \circ d \mathbb{X}^{(a)}_u 
    \end{equation}
where the integral is to be interpreted in the Stratonovich sense.

\begin{example}Let $X:[0, T]\to \mathbb R^2$ be a semimartingale.
\begin{enumerate}
    \item $\mathbb X_{s,t}^{(1)} = X_t^{(1)} - X_s^{(1)}$.
    \item $\mathbb X_{s, t}^{(1, 2)} = \int_s^t X_{s, u}^{(1)} \circ dX_u^{(2)}$.
    \item $\mathbb X_{s, t}^{(2, 2)} = \frac{1}{2}(X_s^{(2)} - X_t^{(2)})^2$.
\end{enumerate}

\end{example}

A more detailed overview of signatures is included in Appendix \ref{sec:overview sigs}.
\end{definition}

\section{Signature Model}
In this section we define the \textit{Signature Model} for asset price dynamics that we propose in this paper. The goal is to approximate the volatility process $\Sigma_t$ (that is a continuous function on the driving Brownian path) by a linear functional on the signature of the Brownian path. 

\begin{definition}[Signature Model]\label{def:sig model}
Let $W$ be a one-dimensional Brownian motion. Let $N\in \NN$ be the \textit{order} of the Signature Model. The Signature Model of parameter $\ell = \{\ell^{(K)} : K \in \mathcal{I}^N_2\}$ is given by $\Sigma_t := \langle \ell, \widehat{\mathbb W}_{0,t}\rangle$, where $\widehat{\mathbb W}$ denotes the signature of $W^{\text{add-time}}$. In other words, the asset price dynamics are given by
\begin{equation}\label{eq:sig model}
dX_t=\langle \ell, \widehat{\mathbb W}_{0,t} \rangle dW_t,\quad X_0=x\in \mathbb R.
\end{equation}
\end{definition}

We note that the Signature Model has two components: the hyperparameter $N\in \NN$, and the model parameter $\ell$. Intuitively, the hyperparameter $N$ plays a similar role to the width of a layer in a neural network. The larger this value is, the richer the range of market dynamics the Signature Models can generate. Once the value of $N$ is fixed, the challenge is to find a suitable model parameter $\ell$. Again, in analogy with neural networks, $\ell$ plays the role of the weights of the network.

The Signature Model possesses the \textit{universality property}, in the sense that given a classical model, there exists a Signature Model that can approximate its dynamics to a given accuracy \cite{levin2013learning}.

We show in the upcoming Sections \ref{sec:simulation}-\ref{sec:calibration} that (a) the Signature Model is efficient to simulate, (b) it is efficient to calibrate, and (c) exotic options can be priced fast under the Signature Model.

\begin{remark}
The Signature Model introduced in Definition \ref{def:sig model} assumes that the source of noise (i.e. the Brownian motion $W$) is one-dimensional. This was done for simplicity, but the authors would like to emphasise that the model generalises in a straightforward way to multi-dimensional Brownian motion.
\end{remark}

\section{Numerical experiments}

We now demonstrate the feasibility of our methodology as outlined in Sections \ref{sec:simulation}-\ref{sec:calibration}. Throughout this section, we work with the Signature Model
$$dX_t=\langle \ell, \widehat{\mathbb W}_{0,t} \rangle dW_t,\quad X_0=1$$
with $\ell=\{\ell^{(K)} : K\in \mathcal I_2^N\}$. We fix $N=4$. Therefore, the model has $1+2+2^2+2^3+2^4=31$ parameters that need to be calibrated. We also fix the terminal maturity $T=1$.

In this section we will show experiments for the calibration of the model, pricing of options under the signature model and simulation. Sections \ref{sec:simulation}-\ref{sec:calibration} will then include the technical details of how calibration, pricing and simulation of signatures model are done.

\subsection{Calibration}
\begin{figure}[h]
  \centering
  \includegraphics[width=\linewidth]{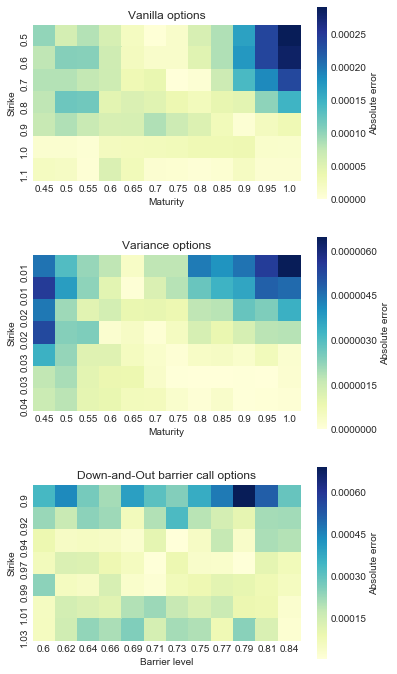}
  \caption{Error analysis between the option prices of the real model and the calibrated Signature Model.}
  \Description{Error analysis.}
  \label{img:calibration}
\end{figure}

We assume that the family of options available on the market are a mixture of vanilla and exotic options, given as follows:

\begin{itemize}
    \item Vanilla call options with strikes $K=0.5, 0.6, \ldots, 1, 1.$ and maturities $t=0.4, 0.45, 0.5, \ldots, 0.9, 0.95, 1$:
    $$\Phi := \max (X_t - K, 0).$$
    \item Variance options with strikes $K=0.01, 0.015, \ldots, 0.035, 0.04$ and maturities $t=0.4, 0.45, 0.5, \ldots, 0.9, 0.95, 1$:
    $$\Phi := \max (\langle X \rangle_t - K, 0).$$
    where $\langle X \rangle$ is the quadratic variation of $X$.
    \sloppy \item Down-and-Out barrier call options with maturity 1, strikes $K=0.9, 0.92, 0.94, \ldots, 1.01, 1.03$ and barrier levels $L=0.6, 0.62, 0.64, \ldots, 0.88, 0.9$:
    $$\Phi := \begin{cases}\max (X_t - K, 0) & \mbox{if }\min_{s \in [0, t]} X_s > L\\ 0 & \mbox{else}. \end{cases}$$
\end{itemize}

The option prices are generated from a Black-Scholes model with volatility $\sigma=0.2$:
$$dX_t = \sigma X_t dW_t.$$

The optimisation \eqref{eq:min} was then solved to calibrate the model parameters $\ell=\{\ell^{(K)} : K\in \mathcal I_2^N\}$.

Figure \ref{img:calibration} shows the absolute error between the real option prices and the option prices of the calibrated model, for the different option types.
\subsection{Simulation}

\begin{figure}[h]
  \centering
  \includegraphics[width=\linewidth]{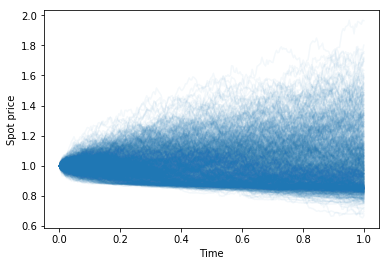}
  \caption{1,000 realisations of the calibrated Signature Model.}
  \Description{Simulation.}
  \label{img:simulation}
\end{figure}

Once the Signature Model has been calibrated to the available option prices, we can use Algorithm \ref{algo:simulation} to simulate realisations of the calibrated Signature Model. Figure \ref{img:simulation} shows 1,000 realisations of the Signature Model.

\subsection{Pricing}

\begin{figure}[h]
  \centering
  \includegraphics[width=\linewidth]{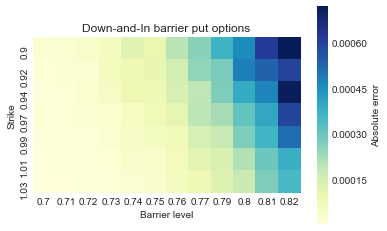}
  \caption{Error analysis between the option prices of the real model and the calibrated Signature Model.}
  \Description{Error analysis.}
  \label{img:pricing}
\end{figure}

We will now use the calibrated Signature Model to price a new set of options that was not used in the calibration step. This set of option consists of Down-and-In barrier put options with barriers levels $L=0.7, 0.71, \ldots, 0.81, 0.82$ and strikes $K=0.9, 0.92, \ldots, 1.01, 1.03$:
$$\Phi := \begin{cases}\max (K - X_t, 0) & \mbox{if }\min_{s \in [0, t]} X_s < L\\ 0 & \mbox{else}. \end{cases}$$

Figure \ref{img:pricing} shows the absolute error of the prices under the Signature Model, compared to the real prices.

As we see, the calibrated model is able to generate accurate prices for these new exotic options. The error is highest when the barrier is close to the strike price, as expected.

\section{Simulation}\label{sec:simulation}

This section will address the question of simulation efficiency of Signature Models. We begin by stating the following two results. The first result rewrites the differential equation \eqref{eq:sig model} solely in terms of the lead-lag signature of the Brownian motion, $\mathbb{\widehat W}_{0, t}^{LL}$. Here $\widehat{\mathbb W}^{LL}$ denotes the lead-lag transformation of $\widehat W$, see Appendix \ref{sec:leadlag}. We use the lead-lag transformation because it allows us to rewrite It\^o integrals as certain Stratonovich integrals, which in turn can be written as linear functions on signatures. The second result guarantees that the computational cost of computing $\mathbb{\widehat W}_{0, t}^{LL}$ is the same as the cost of computing $\{\mathbb{\widehat W}_{0, s}^{LL} \;;\; 0\leq s \leq t\}$. These two results lead to Algorithm \ref{algo:simulation}, which provides an efficient algorithm to sample from a Signature Model.

\begin{proposition}[{\cite[Lemma 3.11]{lyons2019nonparametric}}]\label{prop:LL}
Let $X$ follow a Signature Model with parameter $\ell = \{\ell^{(K)} : K \in \mathcal{I}^N_2\}$. Then, $X$ is given by
\begin{equation}
X_t = \langle x (\o) + \ell \otimes (4), \widehat{\mathbb W}_{0,t}^{LL}\rangle    
\end{equation}

where $\ell \otimes (4) = \{K \otimes (4) : K \in \ell \}$, $x=X_0 \in \mathbb{R}$, and $\widehat{\mathbb W}^{LL}$ denotes the lead-lag transformation, introduced in Definition \ref{def:leadlag}, of the 2-dimensional process $\widehat W=(t, W_t)$.
\end{proposition}

\begin{theorem}[{Chen's identity, \cite[Theorem 2.12]{lyons1998differential}}]\label{th:chen}
Let $0\leq s \leq t$. Then, for each multi-index $K \in \mathcal I_d$ we have
\begin{equation}\label{eq:chen}
\widehat{\mathbb W}_{0,t}^{LL, (K)} =  \sum_{\substack{I, J \in \mathcal I_d \\ I \otimes J = K}} \widehat{\mathbb W}_{0,t}^{LL, (I)} \cdot \widehat{\mathbb W}_{0,t}^{LL, (J)}
\end{equation}
where for any multi-index $K \in \mathcal{I}_d$ we used the notation $\widehat{\mathbb W}_{0,t}^{LL, (K)} = \langle K, \widehat{\mathbb W}_{0,t}^{LL} \rangle$.
\end{theorem}

These two results lead to Algorithm \ref{algo:simulation}. We note there are a number of publicly available software packages to compute signatures, such as \texttt{esig} \footnote{\url{https://pypi.org/project/esig/}}, \texttt{iisignature} \footnote{\url{https://github.com/bottler/iisignature}, \cite{iisignature}} and \texttt{signatory} \footnote{\url{https://github.com/patrick-kidger/signatory}, \cite{signatory}}.

\begin{algorithm}[t]
	\SetAlgoLined
	\SetKwInOut{Parameters}{Parameters}
	\Parameters           {
		$D=\{t_i\}_{i=1}^n$ with {\small$0 = t_0 < t_1 < \ldots <t_{n-1}<t_n=T$: sampling times.}\\
		{\small
		$\ell = \{\ell^{(K)} : K \in \mathcal{I}^N_4\}$: Signature Model parameter.}\\
		$x \in \mathbb R$: initial spot price.		
		 }
	\KwOut{A sample path $\{X_{t_k}\}_{k=0}^n$ from the Signature Model.}
	\BlankLine
	Simulate a one-dimensional Brownian motion at the sampling times $\{W_{t_i}\}_{i=0}^n$.\\
	Apply the lead-lag transformation \eqref{eq:leadlag} to $\widehat W$ to obtain $\widehat W^{LL}$.\\
	{\small
    $\widehat {\mathbb W}^{LL}_{0,0} \leftarrow \{F^{(K)} : K \in \mathcal I_4^{N+1}\}$ with $F^{(\o)}=1$ and $F^{(K)}=0$ for $K \neq \o$.
    }\\
	$X_0 \leftarrow x$.\\
	\For{$k=1, \ldots, n$}{
	    {\small
		Compute the signature $\widehat{\mathbb W}^{LL}_{t_{k-1}, t_k} = \{ \widehat{\mathbb W}_{t_{k-1},t_k}^{LL, (K)}  : K \in \mathcal I_4^{N+1}\}$.
		}\\
		Use Chen's identity (Theorem \ref{th:chen}) to compute the signature $\widehat {\mathbb W}^{LL}_{0,t_k} \leftarrow \{ \widehat{\mathbb W}_{0,t_k}^{LL,K} : K \in \mathcal I_4^{N+1}\}$\\
		Use proposition \ref{prop:LL} to get $X_{t_k} \leftarrow  \langle x (\o) + \ell \otimes (4), \widehat{\mathbb W}_{0,t_k}^{LL}\rangle   $.

	}
	\Return{$\{X_{t_k}\}_{k=0}^n$.}
	\caption{Sampling from a Signature Model.}
	\label{algo:simulation}
\end{algorithm}

\section{Pricing}\label{sec:pricing}

This section will show that exotic options can be priced fast under a Signature Model. This will be done via a two step procedure. First, it was shown in \cite{lyons2019nonparametric, lyons2020numerical} that prices of exotic options can be approximated with arbitrary precision by a special class of payoffs called \textit{signature payoffs}, defined below. Hence, we will assume that the exotic option to be priced is a signature payoff, defined as follows.

\begin{definition}[Signature payoffs]
A signature payoff of maturity $T>0$ and parameter $f = \{f^{(K)} : K \in \mathcal{I}^N_3\}$ is a payoff that pays at time $T$ an amount given by $\langle f, \mathbb{\widehat X}_{0,T}\rangle$.
\end{definition}

Second, the price of a signature payoff is $\langle f, \mathbb E[\mathbb{\widehat{X}}_{0,T}]\rangle$. To price a signature payoff, all we need is $\EE[\mathbb{\widehat{X}}_{0,T}]$, which doesn't depend on the signature payoff itself. In particular, it may be reused to price other signature payoffs.

We now explicitly derive the expected signature $\EE[\mathbb{\widehat{X}}_{0,T}]$ in terms of the model parameters and the expected signature of the lead-lag Brownian motion $\EE[\mathbb{\widehat{W}}^{LL}_{0,T}]$.

\begin{proposition}\label{prop:pricing}
Let $X$ be a Signature Model of order $N\in \NN$ with parameter $\ell = \{\ell^{(K)} : K \in \mathcal{I}^N_2\}$. Consider the following linear functionals $P_1 = (1)$ and $P_2 = \ell \otimes (4)$. Consider any multi-index $I = (i_1, \ldots , i_n) \in \mathcal{I}_2^n$ such that $n \leq N$. Then
\begin{equation}
\mathbb X_{s,t}^{(I)} = \langle C_I(\ell), \widehat{\mathbb W}_{s,t}^{LL} \rangle   
\end{equation}

where $C_I(\ell)$ is given explicitly in closed-form by
\begin{equation}
   C_I(\ell) = (\ldots ((P_{i_1} \succ P_{i_2}) \succ P_{i_3}) \succ \ldots \succ P_{i_n})
\end{equation}
\end{proposition}

\begin{proof}
By Proposition \ref{prop:LL} we know that if $X$ follows a Signature Model with parameter $\ell = \{\ell^{(K)} : K \in \mathcal{I}^N_2\}$ then 
\begin{equation*}
    X_t = \langle x (\o) + \ell \otimes (4), \widehat{\mathbb W}_{0,t}^{LL}\rangle
\end{equation*}

Let $I = (i_1, \ldots, i_n)$ be any multi-index in $\mathcal{I}_2^n$ such that $n\leq N$. If $n=1$ then $I = (i_1)$ and we necessarily one of the following two options must hold

\begin{itemize}
    \item If $i_1 = 1$ then $\mathbb X_{s,t}^{(i_1)} = t - s =  \widehat{\mathbb W}_{s,t}^{LL, (1)} = \langle P_1, \widehat{\mathbb W}_{s,t}^{LL} \rangle$
    \item If $i_1 = 2$ then $\mathbb X_{s,t}^{(i_1)} = X_t - X_s = \langle \ell \otimes (4), \widehat{\mathbb W}_{s,t}^{LL}\rangle = \langle P_2, \widehat{\mathbb W}_{s,t}^{LL} \rangle$.
\end{itemize}

\vspace{0.2cm}

Hence the statement holds for $n=1$. Let's assume by induction that the statement holds for any $1 \leq n \leq N$. We write $I = J \otimes (i_n)$ with $i_n \in \{1,2\}$ and $J = (i_1, \ldots, i_{n-1}) \in \mathcal{I}_2^{n-1}$. Clearly $|(i_n)|, |J| < n$, therefore by induction hypothesis 
\begin{equation*}
    \mathbb X_{s,t}^{(i_n)} = \langle C_{(i_n)}(\ell), \widehat{\mathbb W}_{s,t}^{LL} \rangle = \langle P_{i_n}, \widehat{\mathbb W}_{s,t}^{LL} \rangle
\end{equation*}

and
\begin{align*}
    \mathbb X_{s,t}^{(J)} = \langle C_J(\ell), \widehat{\mathbb W}_{s,t}^{LL} \rangle = \langle (\ldots (P_{i_1} \succ P_{i_2}) \succ \ldots \succ P_{i_{n-1}}), \widehat{\mathbb W}_{s,t}^{LL} \rangle
\end{align*}

By definition of the signature (see \ref{def:sig}) we know that
\begin{align*} 
\mathbb X_{s,t}^{(I)} &= \int_s^t \mathbb X_{s,u}^{(J)} \circ d \mathbb X_u^{(i_n)} \\
&= \int_s^t \langle C_J(\ell), \widehat{\mathbb W}_{s,t}^{LL} \rangle  \circ d \langle C_{(i_n)}(\ell), \widehat{\mathbb W}_{s,t}^{LL} \rangle \\
&= \langle C_J(\ell) \succ C_{(i_n)}(\ell), \widehat{\mathbb W}_{s,t}^{LL} \rangle \\
&= \langle (\ldots (P_{i_1} \succ P_{i_2}) \succ \ldots \succ P_{i_{n-1}}) \succ P_{i_n}, \widehat{\mathbb W}_{s,t}^{LL} \rangle
\end{align*}
which concludes the induction.
\end{proof}

\section{Calibration}\label{sec:calibration}

We will now address the task of calibrating a Signature Model. We assume that the market has a family of options $\{\Phi_i\}_{i=1}^n$ whose market prices $\{p_i\}_{i=1}^n$ are observable. Typically $\{\Phi_i\}_{i=1}^n$ will contain vanilla options, together with some exotic options such as various variance or barrier products. Fix $N\in \NN$ be the order of the Signature Model. The challenge here is to find the model parameter $\ell=\{\ell^{(K)} : K\in \mathcal I_2^N\}$ that best fits the data, in the sense that the prices of $\Phi_i$, under the Signature Model with parameter $\ell$, are approximately given by the observed market prices $p_i$.

Following Section \ref{sec:pricing}, we assume that the options $\Phi_i$ are given by signature options. Therefore, we assume that we can write $\Phi_i$ by
$$\Phi_i = \langle \varphi_i, \widehat{\mathbb X}_{0,T}\rangle,\quad \varphi_i = \{\varphi_i^{(K)} : K\in \mathcal I_2^N\}.$$

The minimisation problem we aim to solve now is the following:
\begin{equation}\label{eq:min}
\min_{\ell = \{\ell^{(K)} : K\in \mathcal I_2^N\}} \sum_{i=1}^n \left (\langle \varphi_i, \mathbb E[\widehat{\mathbb X}_{0, T}]\rangle - p_i \right )^2.
\end{equation}
where $\mathbb E[\widehat{\mathbb X}_{0, T}]$ is the expected signature of the Signature Model with parameter $\ell=\{\ell^{(K)} : K\in \mathcal I_2^N\}$.

By Proposition \ref{prop:pricing}, the price of $\Phi_i$, which is given by $\langle \varphi_i, \mathbb E[\widehat{\mathbb X}_{0, T}]\rangle$, can be written as a polynomial on $\ell^{(K)}$. Hence, the optimisation \eqref{eq:min} is rewritten as a minimisation of a polynomial of variables $\ell^{(K)}$, for $K\in \mathcal I_2^N$.

If the number of parameters $\ell^{(K)}$ is large compared to the number of available option prices, the optimisation problem might be overparametrised and there will be multiple solutions to \eqref{eq:min}. In this case, we are in the robust finance setting where there are multiple equivalent martingale measures that fit to the data. If the number of parameters $\ell^{(K)}$ is small, however, we are in the setting of classical mathematical finance modeling and there will in general be a unique solution to \eqref{eq:min}.


\section{Conclusion}

In this paper we have proposed a new model for asset price dynamics called the \textit{signature model}. This model was develop with the objective of satisfying the following properties:

\begin{enumerate}
    \item Universality.
    \item Efficiency of calibration to vanilla and exotic options.
    \item Fast pricing of vanilla and exotic options.
    \item Efficiency of simulation.
\end{enumerate}

Due to the rich properties of signatures, the signature model satisfies all four properties and is, therefore, capable of generating realistic paths without sacrificing the computational feasibility of calibration, pricing and simulation.

Although this paper has focused on the risk-neutral measure $\mathbb Q$, it can also be used to learn the real-world measure $\mathbb P$. One would first calibrate to the risk-neutral measure $\mathbb Q$ and then learn the drift.

\begin{acks}
This work was supported by The Alan Turing Institute under the EPSRC grant EP/N510129/1.
\end{acks}

\bibliographystyle{ACM-Reference-Format}
\bibliography{sample-base}

\appendix
\section{Overview of signatures}\label{sec:overview sigs}

In this section we state some of the main properties of signatures that are used in this paper.

\begin{definition}[Shuffle of multi-indices]
For any two multi-indices $I, J \in \mathcal{I}_d$ and $1$-dimensional multi-indices $a,b \in \mathcal{I}_d^1 = \{1, \ldots, d\}$ we define the \textit{shuffle product} $\shuffle$ recursively as follows: 
\begin{equation}
\o \shuffle I = I \shuffle \o = I    
\end{equation}
and
\begin{equation}
(I\otimes a) \shuffle (J \otimes b) = ((I\otimes a) \shuffle J) \otimes b + (I \shuffle (J \otimes b)) \otimes a
\end{equation}
\end{definition}

\begin{example}We have the following examples for $\mathcal I_4$:
\begin{enumerate}
    \item $(1, 2) \shuffle (3) = (1, 2, 3) + (1, 3, 2) + (3, 1, 2)$.
    \item $(1, 2) \shuffle (3, 4) = 2\cdot (1, 2, 2, 4) + (1, 2, 4, 2) + (2, 1, 2, 4) + (2, 1, 4, 2) + (2, 4, 1, 2)$.
    \item $(2, 1) \shuffle \o = (2, 1)$.
    \item $\o \shuffle (2, 1) = (2, 1)$.
\end{enumerate}
\end{example}

\begin{proposition}[Shuffle identity]
Let $X:[0, T] \to \mathbb R^d$ be a continuous semimartingale. For any two multi-indices $I,J \in \mathcal{I}_d$ the following identity on the Signature of $X$ holds
\begin{equation}
    \langle I \shuffle J, \mathbb{X}_{s,t} \rangle = \langle I, \mathbb{X}_{s,t} \rangle \cdot \langle J, \mathbb{X}_{s,t} \rangle := \mathbb{X}_{s,t}^{(I)} \cdot \mathbb{X}_{s,t}^{(J)} 
\end{equation}
\end{proposition}

\begin{proof}
Theorem 2.15 in \cite{lyons2007differential}.
\end{proof}

\begin{proposition}[Uniqueness of the Signature]
Let $X:[0, T] \to \mathbb R^d$, $Y:[0, T] \to \mathbb R^d$ be two continuous semimartingales. Then 
\begin{equation}
    \forall t \in [0,T], X_t=Y_t \iff \forall K \in \mathcal{I}_d, \mathbb{X}^{(K)}_{s,t} = \mathbb{Y}^{(K)}_{s,t}
\end{equation}
\end{proposition}

\begin{proof}
See main result in \cite{hambly2010uniqueness}.
\end{proof}

\begin{proposition}[Factorial decay]
Given a semimartingale $X:[0,T] \to \mathbb{R}^d$, for any time interval $[s,t] \subset [0,T]$ and any multi-index $K \in \mathcal{I}_d$ such that $|K|=n$
\begin{equation}
    \big|\mathbb{X}^{(K)}_{s,t}\big| = O\left (\frac{1}{n!}\right )
\end{equation}
\end{proposition}

\begin{proof}
Proposition 2.2 in \cite{lyons2007differential}.
\end{proof}

\begin{definition}
For a given time interval $[0,T]$ we call a continuous, surjective, increasing function $\psi : [0,T] \to [0,T]$ a time-reparametrization.
\end{definition}

\begin{proposition}[Invariance to time-reparametrizations]
Let $X:[0,T] \to \mathbb{R}^d$ be a semimartingale and $\psi: [0,T] \to [0,T]$ be a time-reparametrization. Then the Signature of $X$ has the following invariance property
\begin{equation}
    \mathbb{X}_{s,t} = \mathbb{X}_{\psi(s), \psi(t)} \hspace{0.2cm} \forall s,t \in [0,T] \text{ such that } s<t
\end{equation}
\end{proposition}

\begin{definition}[Half-Shuffle]
Let $F$ and $G$ be any two linear functionals. We define their \textit{half-shuffle product} $\succ$ on $\mathbb X_{s,t}$ as the following (Stratonovich) iterated integral on the real line
\begin{equation}
\langle F \succ G, \mathbb X_{s,t} \rangle  = \int_s^t \langle F, \mathbb X_{s,u} \rangle \circ d \, \langle G, \mathbb X_{s,u} \rangle
\end{equation}
\end{definition}

Let $\mathbb{B}$ be a $2$-dimensional Brownian motion, defined for example on the interval $[0,1]$. Consider two linear functionals $F=\{F^{(K)} : K \in \mathcal{I}_d\}$ and $G=\{G^{(K)} : K \in \mathcal{I}_d\}$ defined as 
\begin{equation}
F^{(K)} = \left\{
    \begin{array}{ll}
        1 & \mbox{if } K = (1,2) \\
        0 & \mbox{otherwise}
    \end{array}
\right.
\end{equation}

and
\begin{equation}
G^{(K)} = \left\{
    \begin{array}{ll}
        1 & \mbox{if } K = (2,1) \\
        0 & \mbox{otherwise}
    \end{array}
\right.
\end{equation}

Then the following quantity
\begin{equation}
    \mathcal{A}_{s,t} = \frac{1}{2} \big\langle F \succ G - G \succ F, \mathbb{B}_{s,t} \big\rangle
\end{equation}

is the \textit{Levy area} of the Brownian motion $\mathbb{B}$ on $[s,t] \subset[0,1]$.

\subsubsection{Expected signature}

We will now define the expected signature of a semimartingale.

\begin{definition}[Expected signature]
Let $X:[0, T]\to \mathbb R^d$ be a continuous semimartingale, and let $\mathbb X_{s,t}=\{\mathbb X_{s, t}^{(K)} \in \mathbb R : K\in \mathcal I_d\}$ be its signature. The expected signature of $X$ is defined by
$$\mathbb E[\mathbb X_{s,t}] := \{\mathbb E[\mathbb X_{s, t}^{(K)}] \in \mathbb R : K\in \mathcal I_d\}.$$
\end{definition}

The expected signature -- i.e. the expectation of the iterated integrals \eqref{eq:sig} -- behaves analogously to the moments of random variables, in the sense that under certain assumptions it characterises the law of the stochastic process:

\begin{theorem}[\cite{chevyrev2016characteristic}]
Let $X:[0, T]\to \mathbb R^d$ be a semimartingale. Then, under certain assumptions (see \cite{chevyrev2016characteristic}) the expected signature $\mathbb E[\mathbb X_{0,T}]$ characterises the law of $X$.
\end{theorem}

\section{Time and Lead-lag transformation}\label{sec:leadlag}

\begin{figure}[h]
  \centering
  \includegraphics[width=.8\linewidth]{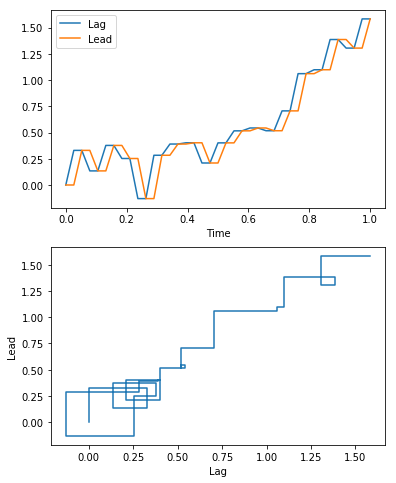}
  \caption{Lead-lag transformation of a Brownian motion.}
  \Description{Lead-lag path.}
  \label{img:ll}
\end{figure}

\sloppy The invariance of the signature of a semimartingale to time reparametrizations allows to handle irregularly sampled sample paths (prices etc.) by completely eliminating the need to retain information about the original time-parametrization. Nonetheless, for the pricing of many options, especially ones resulting from payoffs calculated pathwise (such as integrals for American options), the time represents an important information that we are required to retain. To do so it suffices to augment the state space of the input semimartingale $X$ by adding time $t$ as an extra dimension to get $X^{\text{add-time}}_t = (t, X_t)$. 

We report another basic transformation that can be applied to semimartingales and that will be useful in the sequel of the paper: the lead-lag transformation. This transformation allows us to write It\^o integrals as linear functions on the signature of the lead-lag transformed path.

\begin{definition}[Lead-lag transformation]\label{def:leadlag}
Let $Z:[0, T]\to \mathbb R^d$ be a semimartingale. For each partition $D=\{t_i\}_i\subset [0, T]$ of mesh size $|D|$, define the piecewise linear path $Z^D:[0, T]\to \mathbb R^{2d}$ given by
\begin{align}\label{eq:leadlag}
    &Z_{2kT/2n}^D := (Z_{t_k}, Z_{t_k}),\\
    &Z_{(2k+1)T/2n}^D := (Z_{t_k}, Z_{t_{k+1}})
\end{align}
and linear interpolation in between. Figure \ref{img:ll} shows the lead-lag transformation of a Brownian motion. As we see, the lead component \textit{leads} the lag component, hence the name. The lead component can be seen as the future of the path, and the lag component as the past.

Denote by $\mathbb Z^D$ the signature of $Z^D$. Then, we define the lead-lag transformation of $Z$, denoted by $\mathbb Z^{LL}$, as the limit of signatures of $\mathbb Z^D$:
$$\mathbb Z^{LL} := \lim_{|D|\rightarrow 0} \mathbb Z^D.$$
The work in \cite{flint2016discretely} showed the convergence of this limit and studied some of its properties.
\end{definition}

\subsection{Expected signature of the lead-lag Brownian motion}\label{sec:ES BM}

\begin{definition}
Let $I = (i_1, \ldots ,i_n) \in \mathcal \mathcal{I}^n_3$ be a multi-index. We denote by $\mathcal P(I)$ the set of all possible tuples of non-empty multi-indices from $\mathcal{I}^{n-1}_3$ such that their concatenation is equal to $I$ and their length doesn't exceed $2$, i.e.
\begin{equation*}
\mathcal P(I) = \{(I_1, \ldots, I_k) \in (\mathcal{I}^{n-1}_3)^k : I_1 \otimes \ldots \otimes I_k = I \text{ and } |I_j| \in \{1,2\}\}    
\end{equation*}
\end{definition}

\begin{example}$\phantom{ }$
\begin{enumerate}
\item $\mathcal P((1,2,3)) = \{ (1,2,3), (1, (2,3)), ((1,2), 3)\}$.
\item $\begin{aligned}[t] \mathcal P((1,3,2,2)) =& \{ (1, 3, 2, 2), (1, (3,2), 2), (1, 3, (2,2)), \\ &((1,3),2, 2), ((1,3), (2,2))\}\end{aligned}$
\item $\mathcal P((3,2)) = \{ (3, 2), ((3,2))\}$.
\end{enumerate}
\end{example}

\begin{definition}[Exponential of a linear functional]
Let $F=\{F^{(K)} \in \mathbb{R} : K \in \mathcal{I}_d\}$ be a linear functional. We define the \textit{exponential} of $F$ as the following linear functional
\begin{equation}
    \exp(F) = (\o) + \sum_{k=1}^\infty \frac{1}{k!}F^{\otimes k}
\end{equation}
where for any $k\geq 1, F^{\otimes k} = F\otimes \ldots \otimes F$ for $k$ times.
\end{definition}

\begin{proposition}
Define the function $\alpha: \mathcal{I}_3 \to \mathcal{I}_3$ that maps a multi-index to another multi-index in the following way: $\forall I \in \mathcal{I}_3$
\begin{equation}
    \alpha(I) = \left\{
    \begin{array}{ll}
        (1) & \mbox{if } I = (1) \\
        (2) & \mbox{if } I \in \{(2), (3)\} \\
        -\frac{1}{2}(1) & \mbox{if } I = (2,3) \\
        \frac{1}{2}(1) & \mbox{if } I = (3,2) \\
        0 \cdot (\o) & \mbox{otherwise}
    \end{array}
\right.
\end{equation}
Given a final time $T$ define the linear functional $\mathbf E_T := \exp(T  + \frac{T}{2}(2, 2))$. Then we have the explicit closed-form expression for the Expected Signature of the lead-lag Brownian motion: given any multi-index $I \in \mathcal I_3$
\begin{equation}
\mathbb E\left [\mathbb{\widehat W}_{0,T}^{LL, (I)}\right ] = \sum_{(I_1, \ldots, I_k) \in \mathcal P(I)} \mathbf \langle \alpha(I_1) \otimes \ldots \otimes \alpha(I_k), \mathbf{E}_T \rangle
\end{equation}
\end{proposition}

\begin{proof}
Follows from \cite[Lemma B.1]{lyons2019nonparametric} and the fact that $\mathbb E[\widehat{\mathbb W}_{0, T}] = \mathbf E_T$.
\end{proof}

\begin{example}$\phantom{ }$
If $I = (3,2,3)$, $\mathcal P(I) = \{(3,2,3), ((3,2), 3), (3, (2,3))\}$. Hence,
\begin{align*}
\mathbb E\left [\mathbb{\widehat W}_{0,T}^{LL,(I)}\right ] &= \langle \alpha(3) \otimes \alpha(2) \otimes \alpha(3), \mathbf{E}_T \rangle \\
& \ \ \ + \alpha((3,2)) \otimes \alpha(3), \mathbf{E}_T \rangle \\
& \ \ \ + \alpha(3) \otimes \alpha((2,3)), \mathbf{E}_T \rangle \\
&= \mathbf E_T^{(2,2,2)} + \frac{1}{2} \mathbf E_T^{(1,2)} - \frac{1}{2} \mathbf E_T^{(2,1)}.
\end{align*}
\end{example}

\end{document}

%% file: homemade_macros.tex
\usepackage{amssymb}
\usepackage{mathrsfs}
\usepackage{ifthen}
\usepackage{verbatim}
\usepackage{xcolor}

\usepackage[OT2,OT1]{fontenc}
\newcommand\cyr
{
\renewcommand\rmdefault{wncyr}
\renewcommand\sfdefault{wncyss}
\renewcommand\encodingdefault{OT2}
\normalfont
\selectfont
}
\DeclareTextFontCommand{\textcyr}{\cyr}
\DeclareSymbolFont{cyrletters}{OT2}{wncyss}{m}{n}
\DeclareMathSymbol{\shuffle}{\mathalpha}{cyrletters}{"78}

\usepackage[ruled, linesnumbered]{algorithm2e}
\SetKw{Continue}{continue}
\SetKw{And}{and}
\makeatletter
\newcommand\fs@spaceruled{\def\@fs@cfont{\bfseries}\let\@fs@capt\floatc@ruled
  \def\@fs@pre{\vspace{5\baselineskip}\hrule height.8pt depth0pt \kern2pt}%
  \def\@fs@post{\kern2pt\hrule\relax}%
  \def\@fs@mid{\kern2pt\hrule\kern2pt}%
  \let\@fs@iftopcapt\iftrue}
\makeatother

%% file: mymacros_2.tex
\usepackage{mathrsfs}
\usepackage{ifthen}
\usepackage{verbatim}
\usepackage{fancyhdr}
\usepackage{enumitem}
\usepackage{amsmath,amssymb,amsthm,graphicx}
\usepackage{latexsym}
\usepackage{ifpdf}
\usepackage{appendix}
\usepackage{color}

\theoremstyle{definition}

\def\l{\left}
\def\r{\right}

\newcommand{\ex}[1]{\l[#1\r]}

\renewcommand{\hat}{\widehat}

\newcommand{\EE}[1]{\mathbb{E}\ex{#1}}

\newcommand{\NN}{\mathbb{N}}



%% file: sample-authordraft.bbl

\begin{thebibliography}{34}


\ifx \showCODEN    \undefined \def \showCODEN     #1{\unskip}     \fi
\ifx \showDOI      \undefined \def \showDOI       #1{#1}\fi
\ifx \showISBNx    \undefined \def \showISBNx     #1{\unskip}     \fi
\ifx \showISBNxiii \undefined \def \showISBNxiii  #1{\unskip}     \fi
\ifx \showISSN     \undefined \def \showISSN      #1{\unskip}     \fi
\ifx \showLCCN     \undefined \def \showLCCN      #1{\unskip}     \fi
\ifx \shownote     \undefined \def \shownote      #1{#1}          \fi
\ifx \showarticletitle \undefined \def \showarticletitle #1{#1}   \fi
\ifx \showURL      \undefined \def \showURL       {\relax}        \fi
\providecommand\bibfield[2]{#2}
\providecommand\bibinfo[2]{#2}
\providecommand\natexlab[1]{#1}
\providecommand\showeprint[2][]{arXiv:#2}

\bibitem[\protect\citeauthoryear{Acciaio, Backhoff-Veraguas, and
  Zalashko}{Acciaio et~al\mbox{.}}{2019}]%
        {acciaio2019causal}
\bibfield{author}{\bibinfo{person}{Beatrice Acciaio}, \bibinfo{person}{Julio
  Backhoff-Veraguas}, {and} \bibinfo{person}{Anastasiia Zalashko}.}
  \bibinfo{year}{2019}\natexlab{}.
\newblock \showarticletitle{Causal optimal transport and its links to
  enlargement of filtrations and continuous-time stochastic optimization}.
\newblock \bibinfo{journal}{\emph{Stochastic Processes and their Applications}}
  (\bibinfo{year}{2019}).
\newblock


\bibitem[\protect\citeauthoryear{Boedihardjo, Geng, Lyons, and
  Yang}{Boedihardjo et~al\mbox{.}}{2016}]%
        {boedihardjo2016signature}
\bibfield{author}{\bibinfo{person}{Horatio Boedihardjo}, \bibinfo{person}{Xi
  Geng}, \bibinfo{person}{Terry Lyons}, {and} \bibinfo{person}{Danyu Yang}.}
  \bibinfo{year}{2016}\natexlab{}.
\newblock \showarticletitle{The signature of a rough path: uniqueness}.
\newblock \bibinfo{journal}{\emph{Advances in Mathematics}}
  \bibinfo{volume}{293} (\bibinfo{year}{2016}), \bibinfo{pages}{720--737}.
\newblock


\bibitem[\protect\citeauthoryear{Breeden and Litzenberger}{Breeden and
  Litzenberger}{1978}]%
        {breeden1978prices}
\bibfield{author}{\bibinfo{person}{Douglas~T Breeden} {and}
  \bibinfo{person}{Robert~H Litzenberger}.} \bibinfo{year}{1978}\natexlab{}.
\newblock \showarticletitle{Prices of state-contingent claims implicit in
  option prices}.
\newblock \bibinfo{journal}{\emph{Journal of business}} (\bibinfo{year}{1978}),
  \bibinfo{pages}{621--651}.
\newblock


\bibitem[\protect\citeauthoryear{Cartea, Perez~Arribas, and
  S{\'a}nchez-Betancourt}{Cartea et~al\mbox{.}}{2020}]%
        {cartea2020optimal}
\bibfield{author}{\bibinfo{person}{{\'A}lvaro Cartea}, \bibinfo{person}{Imanol
  Perez~Arribas}, {and} \bibinfo{person}{Leandro S{\'a}nchez-Betancourt}.}
  \bibinfo{year}{2020}\natexlab{}.
\newblock \showarticletitle{Optimal Execution of Foreign Securities: A
  Double-Execution Problem with Signatures and Machine Learning}.
\newblock \bibinfo{journal}{\emph{Available at SSRN}} (\bibinfo{year}{2020}).
\newblock


\bibitem[\protect\citeauthoryear{Chevyrev, Lyons, et~al\mbox{.}}{Chevyrev
  et~al\mbox{.}}{2016}]%
        {chevyrev2016characteristic}
\bibfield{author}{\bibinfo{person}{Ilya Chevyrev}, \bibinfo{person}{Terry
  Lyons}, {et~al\mbox{.}}} \bibinfo{year}{2016}\natexlab{}.
\newblock \showarticletitle{Characteristic functions of measures on geometric
  rough paths}.
\newblock \bibinfo{journal}{\emph{The Annals of Probability}}
  \bibinfo{volume}{44}, \bibinfo{number}{6} (\bibinfo{year}{2016}),
  \bibinfo{pages}{4049--4082}.
\newblock


\bibitem[\protect\citeauthoryear{Chevyrev and Oberhauser}{Chevyrev and
  Oberhauser}{2018}]%
        {chevyrev2018signature}
\bibfield{author}{\bibinfo{person}{Ilya Chevyrev} {and} \bibinfo{person}{Harald
  Oberhauser}.} \bibinfo{year}{2018}\natexlab{}.
\newblock \showarticletitle{Signature moments to characterize laws of
  stochastic processes}.
\newblock \bibinfo{journal}{\emph{arXiv preprint arXiv:1810.10971}}
  (\bibinfo{year}{2018}).
\newblock


\bibitem[\protect\citeauthoryear{Cuchiero, Khosrawi, and Teichmann}{Cuchiero
  et~al\mbox{.}}{2020}]%
        {cuchiero2020generative}
\bibfield{author}{\bibinfo{person}{Christa Cuchiero}, \bibinfo{person}{Wahid
  Khosrawi}, {and} \bibinfo{person}{Josef Teichmann}.}
  \bibinfo{year}{2020}\natexlab{}.
\newblock \showarticletitle{A generative adversarial network approach to
  calibration of local stochastic volatility models}.
\newblock \bibinfo{journal}{\emph{arXiv preprint arXiv:2005.02505}}
  (\bibinfo{year}{2020}).
\newblock


\bibitem[\protect\citeauthoryear{Fermanian}{Fermanian}{2019}]%
        {fermanian2019embedding}
\bibfield{author}{\bibinfo{person}{Adeline Fermanian}.}
  \bibinfo{year}{2019}\natexlab{}.
\newblock \showarticletitle{Embedding and learning with signatures}.
\newblock \bibinfo{journal}{\emph{arXiv preprint arXiv:1911.13211}}
  (\bibinfo{year}{2019}).
\newblock


\bibitem[\protect\citeauthoryear{Flint, Hambly, and Lyons}{Flint
  et~al\mbox{.}}{2016}]%
        {flint2016discretely}
\bibfield{author}{\bibinfo{person}{Guy Flint}, \bibinfo{person}{Ben Hambly},
  {and} \bibinfo{person}{Terry Lyons}.} \bibinfo{year}{2016}\natexlab{}.
\newblock \showarticletitle{Discretely sampled signals and the rough Hoff
  process}.
\newblock \bibinfo{journal}{\emph{Stochastic Processes and their Applications}}
  \bibinfo{volume}{126}, \bibinfo{number}{9} (\bibinfo{year}{2016}),
  \bibinfo{pages}{2593--2614}.
\newblock


\bibitem[\protect\citeauthoryear{Gierjatowicz~P.}{Gierjatowicz~P.}{2020}]%
        {Szpruch2020}
\bibfield{author}{\bibinfo{person}{M.~Siska D. Szpruch~L. Gierjatowicz~P.,
  Sabate-Vidales}.} \bibinfo{year}{2020}\natexlab{}.
\newblock \showarticletitle{Robust Pricing and Hedging with neural SDEs}.
\newblock \bibinfo{journal}{\emph{to appear}} (\bibinfo{year}{2020}).
\newblock


\bibitem[\protect\citeauthoryear{Goodfellow, Pouget-Abadie, Mirza, Xu,
  Warde-Farley, Ozair, Courville, and Bengio}{Goodfellow et~al\mbox{.}}{2014}]%
        {goodfellow2014generative}
\bibfield{author}{\bibinfo{person}{Ian Goodfellow}, \bibinfo{person}{Jean
  Pouget-Abadie}, \bibinfo{person}{Mehdi Mirza}, \bibinfo{person}{Bing Xu},
  \bibinfo{person}{David Warde-Farley}, \bibinfo{person}{Sherjil Ozair},
  \bibinfo{person}{Aaron Courville}, {and} \bibinfo{person}{Yoshua Bengio}.}
  \bibinfo{year}{2014}\natexlab{}.
\newblock \showarticletitle{Generative adversarial nets}. In
  \bibinfo{booktitle}{\emph{Advances in neural information processing
  systems}}. \bibinfo{pages}{2672--2680}.
\newblock


\bibitem[\protect\citeauthoryear{Gyurk{\'o}, Lyons, Kontkowski, and
  Field}{Gyurk{\'o} et~al\mbox{.}}{2013}]%
        {gyurko2013extracting}
\bibfield{author}{\bibinfo{person}{Lajos~Gergely Gyurk{\'o}},
  \bibinfo{person}{Terry Lyons}, \bibinfo{person}{Mark Kontkowski}, {and}
  \bibinfo{person}{Jonathan Field}.} \bibinfo{year}{2013}\natexlab{}.
\newblock \showarticletitle{Extracting information from the signature of a
  financial data stream}.
\newblock \bibinfo{journal}{\emph{arXiv preprint arXiv:1307.7244}}
  (\bibinfo{year}{2013}).
\newblock


\bibitem[\protect\citeauthoryear{Hambly and Lyons}{Hambly and Lyons}{2010}]%
        {hambly2010uniqueness}
\bibfield{author}{\bibinfo{person}{Ben Hambly} {and} \bibinfo{person}{Terry
  Lyons}.} \bibinfo{year}{2010}\natexlab{}.
\newblock \showarticletitle{Uniqueness for the signature of a path of bounded
  variation and the reduced path group}.
\newblock \bibinfo{journal}{\emph{Annals of Mathematics}}
  (\bibinfo{year}{2010}), \bibinfo{pages}{109--167}.
\newblock


\bibitem[\protect\citeauthoryear{Kalsi, Lyons, and Arribas}{Kalsi
  et~al\mbox{.}}{2020}]%
        {kalsi2020optimal}
\bibfield{author}{\bibinfo{person}{Jasdeep Kalsi}, \bibinfo{person}{Terry
  Lyons}, {and} \bibinfo{person}{Imanol~Perez Arribas}.}
  \bibinfo{year}{2020}\natexlab{}.
\newblock \showarticletitle{Optimal execution with rough path signatures}.
\newblock \bibinfo{journal}{\emph{SIAM Journal on Financial Mathematics}}
  \bibinfo{volume}{11}, \bibinfo{number}{2} (\bibinfo{year}{2020}),
  \bibinfo{pages}{470--493}.
\newblock


\bibitem[\protect\citeauthoryear{Karatzas and Shreve}{Karatzas and
  Shreve}{2012}]%
        {karatzas2012brownian}
\bibfield{author}{\bibinfo{person}{Ioannis Karatzas} {and}
  \bibinfo{person}{Steven Shreve}.} \bibinfo{year}{2012}\natexlab{}.
\newblock \bibinfo{booktitle}{\emph{Brownian motion and stochastic calculus.
  Vol. Vol. 113}}.
\newblock \bibinfo{publisher}{Springer Science \& Business Media}.
\newblock


\bibitem[\protect\citeauthoryear{Kidger, Bonnier, Arribas, Salvi, and
  Lyons}{Kidger et~al\mbox{.}}{2019}]%
        {kidger2019deep}
\bibfield{author}{\bibinfo{person}{Patrick Kidger}, \bibinfo{person}{Patric
  Bonnier}, \bibinfo{person}{Imanol~Perez Arribas}, \bibinfo{person}{Cristopher
  Salvi}, {and} \bibinfo{person}{Terry Lyons}.}
  \bibinfo{year}{2019}\natexlab{}.
\newblock \showarticletitle{Deep Signature Transforms}. In
  \bibinfo{booktitle}{\emph{Advances in Neural Information Processing
  Systems}}. \bibinfo{pages}{3099--3109}.
\newblock


\bibitem[\protect\citeauthoryear{Kidger and Lyons}{Kidger and Lyons}{2020}]%
        {signatory}
\bibfield{author}{\bibinfo{person}{Patrick Kidger} {and} \bibinfo{person}{Terry
  Lyons}.} \bibinfo{year}{2020}\natexlab{}.
\newblock \showarticletitle{Signatory: differentiable computations of the
  signature and logsignature transforms, on both CPU and GPU}.
\newblock \bibinfo{journal}{\emph{arXiv preprint arXiv:2001.00706}}
  (\bibinfo{year}{2020}).
\newblock


\bibitem[\protect\citeauthoryear{Kingma and Welling}{Kingma and
  Welling}{2013}]%
        {kingma2013auto}
\bibfield{author}{\bibinfo{person}{Diederik~P Kingma} {and}
  \bibinfo{person}{Max Welling}.} \bibinfo{year}{2013}\natexlab{}.
\newblock \showarticletitle{Auto-encoding variational bayes}.
\newblock \bibinfo{journal}{\emph{arXiv preprint arXiv:1312.6114}}
  (\bibinfo{year}{2013}).
\newblock


\bibitem[\protect\citeauthoryear{Kir{\'a}ly and Oberhauser}{Kir{\'a}ly and
  Oberhauser}{2016}]%
        {kiraly2016kernels}
\bibfield{author}{\bibinfo{person}{Franz~J Kir{\'a}ly} {and}
  \bibinfo{person}{Harald Oberhauser}.} \bibinfo{year}{2016}\natexlab{}.
\newblock \showarticletitle{Kernels for sequentially ordered data}.
\newblock \bibinfo{journal}{\emph{arXiv preprint arXiv:1601.08169}}
  (\bibinfo{year}{2016}).
\newblock


\bibitem[\protect\citeauthoryear{Levin, Lyons, and Ni}{Levin
  et~al\mbox{.}}{2013}]%
        {levin2013learning}
\bibfield{author}{\bibinfo{person}{Daniel Levin}, \bibinfo{person}{Terry
  Lyons}, {and} \bibinfo{person}{Hao Ni}.} \bibinfo{year}{2013}\natexlab{}.
\newblock \showarticletitle{Learning from the past, predicting the statistics
  for the future, learning an evolving system}.
\newblock \bibinfo{journal}{\emph{arXiv preprint arXiv:1309.0260}}
  (\bibinfo{year}{2013}).
\newblock


\bibitem[\protect\citeauthoryear{Li, Zhang, and Jin}{Li et~al\mbox{.}}{2017}]%
        {li2017lpsnet}
\bibfield{author}{\bibinfo{person}{Chenyang Li}, \bibinfo{person}{Xin Zhang},
  {and} \bibinfo{person}{Lianwen Jin}.} \bibinfo{year}{2017}\natexlab{}.
\newblock \showarticletitle{Lpsnet: A novel log path signature feature based
  hand gesture recognition framework}. In \bibinfo{booktitle}{\emph{Proceedings
  of the IEEE International Conference on Computer Vision Workshops}}.
  \bibinfo{pages}{631--639}.
\newblock


\bibitem[\protect\citeauthoryear{Lyons, Nejad, and Arribas}{Lyons
  et~al\mbox{.}}{2019}]%
        {lyons2019nonparametric}
\bibfield{author}{\bibinfo{person}{Terry Lyons}, \bibinfo{person}{Sina Nejad},
  {and} \bibinfo{person}{Imanol~Perez Arribas}.}
  \bibinfo{year}{2019}\natexlab{}.
\newblock \showarticletitle{Nonparametric pricing and hedging of exotic
  derivatives}.
\newblock \bibinfo{journal}{\emph{arXiv preprint arXiv:1905.00711}}
  (\bibinfo{year}{2019}).
\newblock


\bibitem[\protect\citeauthoryear{Lyons, Nejad, and Perez~Arribas}{Lyons
  et~al\mbox{.}}{2020}]%
        {lyons2020numerical}
\bibfield{author}{\bibinfo{person}{Terry Lyons}, \bibinfo{person}{Sina Nejad},
  {and} \bibinfo{person}{Imanol Perez~Arribas}.}
  \bibinfo{year}{2020}\natexlab{}.
\newblock \showarticletitle{Numerical Method for Model-free Pricing of Exotic
  Derivatives in Discrete Time Using Rough Path Signatures}.
\newblock \bibinfo{journal}{\emph{Applied Mathematical Finance}}
  (\bibinfo{year}{2020}), \bibinfo{pages}{1--15}.
\newblock


\bibitem[\protect\citeauthoryear{Lyons, Ni, and Oberhauser}{Lyons
  et~al\mbox{.}}{2014}]%
        {lyons2014feature}
\bibfield{author}{\bibinfo{person}{Terry Lyons}, \bibinfo{person}{Hao Ni},
  {and} \bibinfo{person}{Harald Oberhauser}.} \bibinfo{year}{2014}\natexlab{}.
\newblock \showarticletitle{A feature set for streams and an application to
  high-frequency financial tick data}. In \bibinfo{booktitle}{\emph{Proceedings
  of the 2014 International Conference on Big Data Science and Computing}}.
  \bibinfo{pages}{1--8}.
\newblock


\bibitem[\protect\citeauthoryear{Lyons}{Lyons}{1998}]%
        {lyons1998differential}
\bibfield{author}{\bibinfo{person}{Terry~J Lyons}.}
  \bibinfo{year}{1998}\natexlab{}.
\newblock \showarticletitle{Differential equations driven by rough signals}.
\newblock \bibinfo{journal}{\emph{Revista Matem{\'a}tica Iberoamericana}}
  \bibinfo{volume}{14}, \bibinfo{number}{2} (\bibinfo{year}{1998}),
  \bibinfo{pages}{215--310}.
\newblock


\bibitem[\protect\citeauthoryear{Lyons, Caruana, and L{\'e}vy}{Lyons
  et~al\mbox{.}}{2007}]%
        {lyons2007differential}
\bibfield{author}{\bibinfo{person}{Terry~J Lyons}, \bibinfo{person}{Michael
  Caruana}, {and} \bibinfo{person}{Thierry L{\'e}vy}.}
  \bibinfo{year}{2007}\natexlab{}.
\newblock \bibinfo{booktitle}{\emph{Differential equations driven by rough
  paths}}.
\newblock \bibinfo{publisher}{Springer}.
\newblock


\bibitem[\protect\citeauthoryear{Reizenstein and Graham}{Reizenstein and
  Graham}{2020}]%
        {iisignature}
\bibfield{author}{\bibinfo{person}{Jeremy~F Reizenstein} {and}
  \bibinfo{person}{Benjamin Graham}.} \bibinfo{year}{2020}\natexlab{}.
\newblock \showarticletitle{Algorithm 1004: The iisignature library: Efficient
  calculation of iterated-integral signatures and log signatures}.
\newblock \bibinfo{journal}{\emph{ACM Transactions on Mathematical Software
  (TOMS)}} \bibinfo{volume}{46}, \bibinfo{number}{1} (\bibinfo{year}{2020}),
  \bibinfo{pages}{1--21}.
\newblock


\bibitem[\protect\citeauthoryear{Vapnik}{Vapnik}{2013}]%
        {vapnik2013nature}
\bibfield{author}{\bibinfo{person}{Vladimir Vapnik}.}
  \bibinfo{year}{2013}\natexlab{}.
\newblock \bibinfo{booktitle}{\emph{The nature of statistical learning
  theory}}.
\newblock \bibinfo{publisher}{Springer science \& business media}.
\newblock


\bibitem[\protect\citeauthoryear{Xie, Sun, Jin, Ni, and Lyons}{Xie
  et~al\mbox{.}}{2017}]%
        {xie2017learning}
\bibfield{author}{\bibinfo{person}{Zecheng Xie}, \bibinfo{person}{Zenghui Sun},
  \bibinfo{person}{Lianwen Jin}, \bibinfo{person}{Hao Ni}, {and}
  \bibinfo{person}{Terry Lyons}.} \bibinfo{year}{2017}\natexlab{}.
\newblock \showarticletitle{Learning spatial-semantic context with fully
  convolutional recurrent network for online handwritten chinese text
  recognition}.
\newblock \bibinfo{journal}{\emph{IEEE transactions on pattern analysis and
  machine intelligence}} \bibinfo{volume}{40}, \bibinfo{number}{8}
  (\bibinfo{year}{2017}), \bibinfo{pages}{1903--1917}.
\newblock


\bibitem[\protect\citeauthoryear{Yang, Jin, and Liu}{Yang
  et~al\mbox{.}}{2015}]%
        {yang2015chinese}
\bibfield{author}{\bibinfo{person}{Weixin Yang}, \bibinfo{person}{Lianwen Jin},
  {and} \bibinfo{person}{Manfei Liu}.} \bibinfo{year}{2015}\natexlab{}.
\newblock \showarticletitle{Chinese character-level writer identification using
  path signature feature, DropStroke and deep CNN}. In
  \bibinfo{booktitle}{\emph{2015 13th International Conference on Document
  Analysis and Recognition (ICDAR)}}. IEEE, \bibinfo{pages}{546--550}.
\newblock


\bibitem[\protect\citeauthoryear{Yang, Jin, and Liu}{Yang
  et~al\mbox{.}}{2016a}]%
        {yang2016deepwriterid}
\bibfield{author}{\bibinfo{person}{Weixin Yang}, \bibinfo{person}{Lianwen Jin},
  {and} \bibinfo{person}{Manfei Liu}.} \bibinfo{year}{2016}\natexlab{a}.
\newblock \showarticletitle{Deepwriterid: An end-to-end online text-independent
  writer identification system}.
\newblock \bibinfo{journal}{\emph{IEEE Intelligent Systems}}
  \bibinfo{volume}{31}, \bibinfo{number}{2} (\bibinfo{year}{2016}),
  \bibinfo{pages}{45--53}.
\newblock


\bibitem[\protect\citeauthoryear{Yang, Jin, Ni, and Lyons}{Yang
  et~al\mbox{.}}{2016b}]%
        {yang2016rotation}
\bibfield{author}{\bibinfo{person}{Weixin Yang}, \bibinfo{person}{Lianwen Jin},
  \bibinfo{person}{Hao Ni}, {and} \bibinfo{person}{Terry Lyons}.}
  \bibinfo{year}{2016}\natexlab{b}.
\newblock \showarticletitle{Rotation-free online handwritten character
  recognition using dyadic path signature features, hanging normalization, and
  deep neural network}. In \bibinfo{booktitle}{\emph{2016 23rd International
  Conference on Pattern Recognition (ICPR)}}. IEEE,
  \bibinfo{pages}{4083--4088}.
\newblock


\bibitem[\protect\citeauthoryear{Yang, Jin, Tao, Xie, and Feng}{Yang
  et~al\mbox{.}}{2016c}]%
        {yang2016dropsample}
\bibfield{author}{\bibinfo{person}{Weixin Yang}, \bibinfo{person}{Lianwen Jin},
  \bibinfo{person}{Dacheng Tao}, \bibinfo{person}{Zecheng Xie}, {and}
  \bibinfo{person}{Ziyong Feng}.} \bibinfo{year}{2016}\natexlab{c}.
\newblock \showarticletitle{DropSample: A new training method to enhance deep
  convolutional neural networks for large-scale unconstrained handwritten
  Chinese character recognition}.
\newblock \bibinfo{journal}{\emph{Pattern Recognition}}  \bibinfo{volume}{58}
  (\bibinfo{year}{2016}), \bibinfo{pages}{190--203}.
\newblock


\bibitem[\protect\citeauthoryear{Yang, Lyons, Ni, Schmid, Jin, and Chang}{Yang
  et~al\mbox{.}}{2017}]%
        {yang2017leveraging}
\bibfield{author}{\bibinfo{person}{Weixin Yang}, \bibinfo{person}{Terry Lyons},
  \bibinfo{person}{Hao Ni}, \bibinfo{person}{Cordelia Schmid},
  \bibinfo{person}{Lianwen Jin}, {and} \bibinfo{person}{Jiawei Chang}.}
  \bibinfo{year}{2017}\natexlab{}.
\newblock \showarticletitle{Leveraging the path signature for skeleton-based
  human action recognition}.
\newblock \bibinfo{journal}{\emph{arXiv preprint arXiv:1707.03993}}
  (\bibinfo{year}{2017}).
\newblock


\end{thebibliography}
